\documentclass[aps,prl,twocolumn,showpacs,superscriptaddress,amsmath,amssymb]{revtex4-1}

\usepackage{graphicx}
\usepackage{hyperref}

\newcommand{\ie}{i.e.\@}
\newcommand{\eg}{e.g.\@}

\newcommand{\eq}[1]{Eq.~\eqref{eq:#1}}
\newcommand{\Eq}[1]{Equation \eqref{eq:#1}}
\newcommand{\fig}[1]{Fig.~\ref{fig:#1}}

\hypersetup{bookmarksnumbered=true,
  bookmarksopen=true,
  citecolor=blue,
  colorlinks=true,
  hypertexnames=true,
  linkcolor=blue,
  linktocpage=true,
  pdfauthor={Matthias Liertzer, Li Ge, Hakan T\"ureci, Stefan Rotter},
  pdfhighlight=/I,
  pdfpagemode=UseNone,
  pdftitle={Pump-induced Exceptional Points in Lasers above
    Threshold},
  urlcolor=blue}

\begin{document}

\title{Pump-induced Exceptional Points in Lasers}

\author{M. Liertzer}
\email{matthias.liertzer@tuwien.ac.at}
\affiliation{ Institute for Theoretical Physics, Vienna University of
              Technology, A-1040 Vienna, Austria, EU}

\author{Li Ge}
\affiliation{ Department of Electrical Engineering, Princeton University,
              Princeton, New Jersey 08544, USA}

\author{A. Cerjan}
\affiliation{Department of Applied Physics, Yale University, New
  Haven, Connecticut 06520, USA}

\author{A. D. Stone}
\affiliation{Department of Applied Physics, Yale University, New
  Haven, Connecticut 06520, USA}

\author{H. E. T\"ureci}
\affiliation{ Department of Electrical Engineering, Princeton University,
              Princeton, New Jersey 08544, USA}
\affiliation{ Institute for Quantum Electronics, ETH-Z\"urich,
              CH-8093 Z\"urich, Switzerland }

\author{S. Rotter}
\email{stefan.rotter@tuwien.ac.at}
\affiliation{ Institute for Theoretical Physics, Vienna University of
              Technology, A-1040 Vienna, Austria, EU}
\date{\today}

\begin{abstract}
  We demonstrate that the above-threshold behavior of a laser can be
  strongly affected by exceptional points which are induced by pumping
  the laser nonuniformly. At these singularities, the eigenstates of
  the non-Hermitian operator which describes the lasing modes
  coalesce. In their vicinity, the laser may turn off even when the
  overall pump power deposited in the system is increased. Such
  signatures of a pump- induced exceptional point can be
  experimentally probed with coupled ridge or microdisk lasers.
\end{abstract}


\maketitle

The interest in physical systems described by non-Hermitian operators
has recently been revived by a number of pioneering experiments
\cite{GuoSal2009,RutMak2010,WanChoGe2011,DieHarKir2011,LeeYanMoo2009,ChoKanLim2010,SchLiZhe2011}
which demonstrated the rich physics induced by the system's
non-Hermiticity. In this context exceptional points (EPs), at which
non-Hermitian operators are defective and pairs of their eigenstates
coalesce \cite{Hei2004}, are of particular interest. In the vicinity
of these EPs the eigenvalues display an intricate topology of
intersecting Riemann sheets, leading to nontrivial geometric phases
when encircling the EP parametrically, as well as to problems with
mode labeling
\cite{DieHarKir2011,LeeYanMoo2009,RyuLeeKim2009,ChoKanLim2010,SchLiZhe2011,WieEbeShi2011,LefAtaSin2009,
  Hei2004}.

Photonic structures in the presence of gain or loss are a natural
arena in which EPs can play a role, since they are described by a
non-Hermitian Maxwell operator, arising from a complex dielectric
function.  Striking examples which have received much attention
recently are optical systems with balanced gain and loss, for which
this operator is invariant under parity ($\mathcal P$) and
time-reversal ($\mathcal T$) \cite{BenBoe1998}.  In such systems, as
the real wave vector, $k$, of the input radiation is varied, the
electromagnetic scattering matrix undergoes a transition at an EP from
a ``flux-conserving'' behavior to an amplifying/attenuating behavior
\cite{ChoGeSto2011}. This transition leads to a number of novel
optical properties \cite{GuoSal2009,RutMak2010}, such as power
oscillations of light, beam steering, and self-sustained radiation
\cite{MakGanChr2008,KlaGueMoi2008,Sch2010}.

Lasers are prototype non-Hermitian systems, with the additional
complexity of being described by a nonlinear wave equation above their
first oscillation threshold. Until very recently, however, no
connection had been made between pumping a laser and the phenomena
associated with EPs. The reason for this, as we shall see below, is
that the non-Hermitian operators describing lasers do not reveal EPs
when the laser is pumped {\it uniformly} in space. Two previous works
have hinted at EP physics in non-uniformly pumped lasers. First, a
work on $\mathcal{PT}$ optical systems has shown that a cavity with
balanced regions of strong gain and strong loss can actually lase {\it
  after} the $\mathcal{ PT}$-breaking EP has been crossed
\cite{Lon2010,ChoGeSto2011}. However, the properties of this unusual
laser (which has yet to be demonstrated experimentally) are not
influenced by the EP {\it above} threshold.  Another laser system
which exhibits EPs is composed of very-low \emph{Q} cavities, which
can form novel surface-state lasing modes \cite{GeChoRot2011}.  Again,
however, these modes are created well away from the EP and their
lasing properties are not affected by it, nor is this a practical
system for experiments.

In the current Letter, we propose a class of non-uniformly pumped
laser systems, which are achievable in practical laser structures and
which show the dramatic role of an EP {\em above} the laser threshold
in controlling and even switching off the coherent emission. The
systems we study consist of two coherently coupled laser cavities as
realized, \eg, by two coupled ridge lasers (see insets in
\begin{figure}
  \centering
  \includegraphics{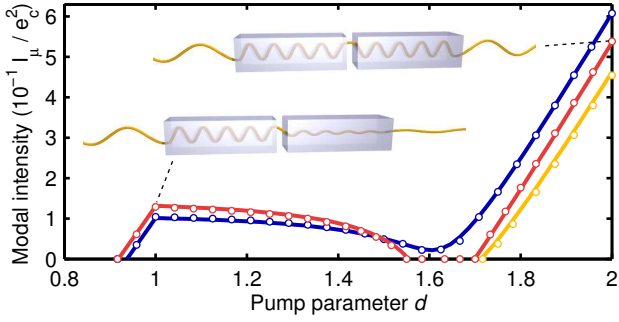}
  \caption{(color online) Intensity output of a laser system
    consisting of two 1D coupled ridge lasers, each of length
    $100\,{\mu}$m with an air gap of size $10\,{\mu}$m and an
    (unpumped) index of refraction $n = 3+0.13i$.  For
    $0\!<\!d\!<\!1$, the pump in the left ridge is linearly increased
    in the range $0\!<\!D\!<\!1.2$, and for $1\!<\!d\!<\!2$, the same
    is done in the right ridge while keeping the left ridge at
    $D=1.2$. The points show the results determined from a
    finite-difference time-domain solution of the Maxwell-Bloch
    equations (using $\gamma_\parallel=0.02\,$mm${}^{-1}$). The solid lines are the
    solutions as given by SALT (see the text). The three different colors
    (red, blue and yellow \cite{Note1}) represent the results for different gain
    curves with a width of $2\gamma_\bot=4\,$mm${}^{-1}$ centered around
    $k_a=\{92.4,94.6,96.3\}\,$mm${}^{-1}$, respectively. Although the
    overall pump power increases monotonically over the entire range
    of $d$, the laser is found to turn off in the range
    $1.55\!<\!d\!<\!1.68$ due to the influence of an exceptional
    point. The insets show the electric field amplitude in the laser
    at the pump steps $d=\{1,2\}$ for the gain curve centered around
    $k_a=94.6\, $mm${}^{-1}$.}
  \label{fig:1}
\end{figure}
\fig{1}).
The cavities are assumed to have large loss in the absence of
pumping. Their unusual behavior when pumped is shown in \fig{1} (red
curve) \footnote{When Figs. \ref{fig:1} and \ref{fig:2} are printed in
  grayscale, note that yellow corresponds to light gray, red to gray,
  and blue to dark gray.}. First the left cavity is pumped
$(0\!<\!d\!<1)$ until it begins lasing in an asymmetric mode which
emits predominantly to the left; then, pumping is added to the right
cavity and increased $(1\!<\!d\!<2)$, which, surprisingly, causes the
laser first to shut off and then to turn on again as a symmetric mode
(details of the pumping scheme are given in the caption).  This
behavior occurs for a specific value of the center and width of the
gain curve.  Quite different behavior occurs if the gain center is
shifted by a small amount compared to the free spectral range.  If
shifted down (yellow curve), the system never lases at all in the
asymmetric mode; if shifted up (blue curve), it does not ever switch
off, but the mode slowly evolves from asymmetric to symmetric with a
strong intensity dip after the initial turn-on. We will show that this
counter-intuitive dependence on the pump strength and high sensitivity
to the gain curve is completely controlled by an EP in the relevant
non-Hermitian operator for the laser problem.

The above results have been obtained from the stationary solutions of
the nonlinear Maxwell-Bloch equations, which are the key equations in
semiclassical laser theory \cite{Hak1985,SarScuLam1977}. To calculate
these solutions, we employed the newly developed steady-state \emph{ab
  initio} laser theory (SALT)
\cite{GeChoSto2010,TurStoRot2009,TurStoCol2006}, which is exact in the
single-mode regime of interest here (except for the standard rotating
wave approximation). Because of the novelty of the behavior studied
here, we independently confirmed the SALT results (see lines in
\fig{1}) using the more familiar finite-difference time-domain method
(see points in \fig{1}) \cite{GeTanSto2008}.  The observed reentrant
behavior, however, cannot be understood in the time domain, requiring,
instead, a frequency-domain approach like SALT. In 1D and 2D, the SALT
equations are coupled scalar nonlinear wave equations for the lasing
modes $\Psi_\mu({\mathbf x})$ and their corresponding lasing
frequencies $k_\mu$, of the form \cite{GeChoSto2010}
\begin{equation}
  \biggl[ \nabla^2 + k_\mu^2 \,\varepsilon\left(\mathbf{x},\{k_\nu,
    \Psi_\nu\}\right)\biggr] \Psi_\mu(\mathbf{x}) = 0\,.
\label{eq:helmholtz}
\end{equation}
Here the complex dielectric function $\varepsilon(\mathbf x, \{k_\nu,
\Psi_\nu\})= \varepsilon_c(\mathbf
x)+\varepsilon_g(\mathbf{x},\{k_\nu, \Psi_\nu\})$ comprises a
contribution from the passive cavity $\varepsilon_c$ as well as a gain
contribution $\varepsilon_g$ induced by the external pump. This active
part
\begin{equation}
  \varepsilon_g\left(\mathbf{x},\{k_\nu, \Psi_\nu\}\right) =
  \frac{\gamma_\bot}{k_\mu -
    k_a + i \gamma_\bot}\frac{D_0(\mathbf{x},d)}{1
    + \sum_\nu \Gamma_\nu |\Psi_\nu|^2}\,,\label{eq:salt_effdielec_gain}
\end{equation}
contains modal interactions and gain saturation, due to spatial hole
burning through the nonlinearity given by $[1 + \sum_\nu \Gamma_\nu
|\Psi_\nu|^2]^{-1}$, where $\Gamma_\nu = \Gamma(k_\nu)$ is a
Lorentzian function of width $2\gamma_\perp$, centered at the atomic
transition frequency $k_a$ and evaluated at the lasing frequency
$k_\mu\,(c=1)$.  Crucially, $\varepsilon_g$ in
\eq{salt_effdielec_gain} also contains the spatially varying inversion
density $D_0(\mathbf{x},d)$, the ``pump'' \footnote{Both the scalar
  mode $\Psi_\mu$, the real part of which corresponds to the electric
  field, and the pump $D_0$ are given here in their natural units
  $e_c$ and $d_c$ \cite{GeChoSto2010}.}, where we have introduced an
adiabatic parameter $d$. This leads us to the concept of a ``pump
trajectory" (see, \eg, \fig{1}) consisting of a sequence of adiabatic
variations in both the pump profile and amplitude, parametrized by $d$
and satisfying $\frac{d D_0}{d t} \ll \gamma_\parallel$ to ensure
adiabaticity ($\gamma_\parallel$ is the longitudinal relaxation rate)
\cite{Hak1985,GeTanSto2008}. We obtain the solutions to \eq{helmholtz}
along such a pump trajectory for both the laser frequency $k_\mu$ and
the corresponding electric field $\Psi_\mu$ as the fixed points of a
self-consistent iteration carried out for each given value of the pump
$D_0(\mathbf{x}, d)$ \cite{Lie2011}.  With this procedure, we find the
full {\it nonlinear} solutions for the emission intensities above
threshold; as we now demonstrate, the qualitative behavior shown in
\fig{1} is, however, determined already by the {\it linear} threshold
conditions for a given pump trajectory.

To show this, we introduce a generalization of the threshold equations
of SALT, which are posed in two steps.  First we define a non-Hermitian
eigenvalue problem,
\begin{equation}
\biggl[ \nabla^2 + k^2 \biggl( \varepsilon_c(k) +
\eta_n(k,d) D_0(\mathbf{x},d)
 \biggr)
\biggr] u_n(k,d;\mathbf{x}) = 0\,,
\label{eq:gtcf}
\end{equation}
where $\eta_n(k,d)$ is the complex eigenvalue
\cite{GeChoSto2010,Lie2011}, and $\eta_n(k,d) D_0(\mathbf{x},d)$ plays
the role of $\varepsilon_g$ in \eq{salt_effdielec_gain} at threshold,
where the nonlinearity vanishes. \Eq{gtcf}, which is solved using
non-Hermitian ``constant flux'' boundary conditions at the laser
boundary ($\partial_x u_n=\pm iku_n$), is a generalization of the
threshold constant flux (TCF) equation \cite{GeChoSto2010} for pump
trajectories; the $\{u_n,\eta_n \}$ are the TCF states and
eigenvalues.

The first laser mode is determined out of the countably infinite set
of ${\eta_n(k,d)}$ as the one which satisfies
\begin{equation}
\eta_{n}(k_\mu,d_\mu) = \frac{\gamma_\bot}{k_\mu - k_a + i\gamma_\bot}\,
\label{eq:tlm_cond}
\end{equation}
for the smallest value of $d_\mu$, which, from \eq{tlm_cond}, also
yields the threshold lasing frequency $k_\mu$. The lasing mode
$\psi_\mu(\mathbf{x})$ at the threshold is then just given by the
corresponding eigenstate $u_{n}(k_\mu,d_\mu;\mathbf{x})$. The
linear non-Hermitian TCF equation, \eq{gtcf}, can, however, be solved
for arbitrary values of $k$ and $d$. We are thus naturally provided with the
two continuous parameters necessary to demonstrate the presence of EPs
in the complex spectrum of the TCF operator \cite{Hei2004}. This will
bring us to the central insight that the TCF operator which determines
the lasing thresholds contains the EPs which control the lasing
behavior shown in \fig{1}.

\begin{figure}[b]
  \centering
  \includegraphics{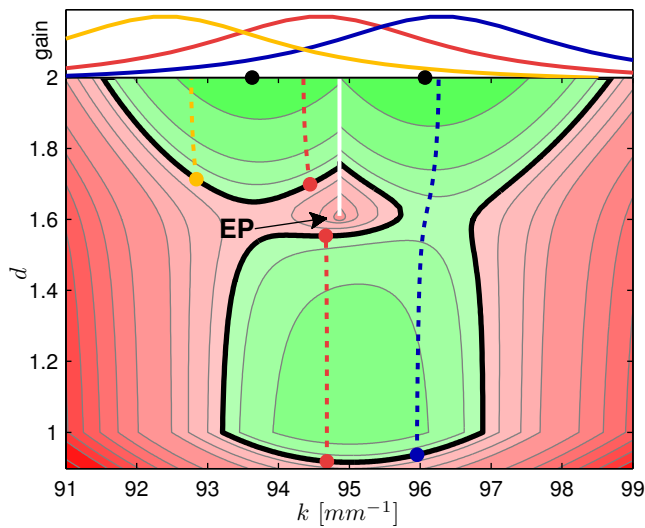}
  \caption{(color online) Contour plot of the function $f(k,d)$, which
    indicates the parameter regions of $k$ and $d$ where the laser in
    \fig{1} is above threshold (green, inside black contour) or below
    threshold (red, outside black contour). The threshold condition,
    $f(k,d)=0$, is satisfied at the solid black contour. The EP in the
    center of the plot pulls the laser below the threshold in its
    vicinity. The frequency dependence of the solutions of the
    nonlinear SALT equations (dashed lines) and the corresponding
    laser thresholds (dots on the solid black contour) are provided
    for the same gain curves (shown in the top panel) as in \fig{1}.
    The evaluation of $f(k,d)$ involves two interconnected Riemann
    sheets (see \fig{3}), resulting in a cut between the two sheets
    (see the white line right above the EP). The two black dots at the
    upper edge of the plot represent the resonance frequencies of the
    passive cavity system.}
  \label{fig:2}
\end{figure}
 
To show this explicitly, we rewrite the complex \eq{tlm_cond} as two
real conditions which neatly separate the role of the TCF spectrum and
the gain curve
\begin{equation}
|\eta_n|^2+\text{Im}(\eta_n) = 0,\quad\qquad \frac{k-k_a}{\gamma_\bot} = \frac{\text{Im}(\eta_n) +
    1}{\text{Re}(\eta_n)}\label{eq:tlm_cond2}\,.
\end{equation}
The left equation has not been analyzed in previous work on SALT; it
has the remarkable property of being entirely independent of the gain
curve parameters $k_a$ and $\gamma_\bot$ (which do not appear in the
TCF equation). Hence, it defines a threshold boundary in the $k,d$
plane dividing lasing and nonlasing regions, solely from the values of
$\{\eta_n\}$. To conveniently capture the entire $k,d$ landscape of
possible lasing solutions, we define a function $f(k,d) \equiv
\min_n[\,|\eta_n|^2+\text{Im}(\eta_n)\,]$, which has the property that
the contour $f(k,d)=0$ is the locus of all possible lasing
thresholds. The actual lasing thresholds on this contour are then
determined by the gain curve parameters in the right-hand condition of
\eq{tlm_cond2}. In \fig{2}, we show the relevant regions of $f(k,d)$
above ($f<0$, green, within black contour) and below ($f>0$, red,
outside black contour) threshold, as well as the threshold boundary
($f=0$, black contour). When $f<0$, the TCF equation does not give the
lasing solutions and full nonlinear SALT is needed to find the
intensity and frequency, as was done for the curves in Figs.~\ref{fig:1} and
\ref{fig:2}.
 
The most striking feature of the contour plot shown in \fig{2} is the
existence of a substantial inclusion of subthreshold region (red,
outside black contour) in the midst of the superthreshold region
(green, within black contour). At the center of this inclusion is a
local maximum of $f$; this point corresponds to an EP of the TCF
operator.  For the middle gain curve (red), there are three threshold
solutions, two very near the EP.  If we order these solutions
according to increasing pump, $d$, we find that $\partial f/\partial d
< 0$ at the first and third thresholds and $\partial f/\partial d > 0$
at the second. As a function of $d$, the laser thus turns on-off-on,
just as found in \fig{1}. Hence, nonlinear effects play no role in the
qualitative behavior of this laser; they only determine the amplitude
of the lasing emission in the green region. All the interesting
behavior is controlled by the EP: It completely suppresses lasing in
its vicinity of $k$ and $d$-parameters, thereby causing the reentrant
lasing (red curve) as well as the nonmonotonic but continuous lasing
emission (blue curve), which is influenced by its proximity to the EP.

In the entire $k$ and $d$ range shown in \fig{2}, exactly only two
eigenvalues $\eta_{1,2}$ contribute to $f(k,d)$. In the coupled
cavities studied here, these two eigenvalues are associated with a
symmetric-antisymmetric doublet of passive cavity resonances (see
black dots in \fig{2}). The doublets are separated from each other by
approximately the free spectral range of a single cavity. If both
cavities are \emph{uniformly} pumped with the same gain curves as in
\fig{2}, the laser emits in modes each of which can be associated with
one of the passive resonances (not shown). However, no EP and hence no
pump-induced inclusion or nonmonotonic laser emission appears.  Only
when nonuniform pumping is applied may a pair of eigenvalues coalesce
at an EP.

\begin{figure}[b]
  \centering
  \includegraphics{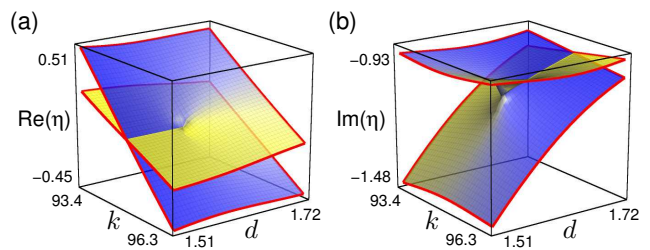}
  \caption{(color online) (a) Real and (b) imaginary parts of the two
    eigenvalues $\eta_{1,2}(k,d)$ which are closest to the threshold
    in the $k$ and $d$ parameter region around the laser turn-off observed
    in \fig{1}. The eigenvalue surfaces display the typical structure
    of intersecting Riemann surfaces, centered around an EP.}
  \label{fig:3}
\end{figure}

In \fig{3}, we plot the $k$ and $d$ dependence of the real and
imaginary part of the two eigenvalues $\eta_{1,2}$. Focusing on the
parameter region around $d=1.55$, where the laser turns off, we find
that the parametric dependence of the eigenvalues on $k$ and $d$ shows
the typical topological structure of two intersecting Riemann sheets
around an EP located at $k\approx95\,mm^{-1}, d\approx1.6$.  Away from
the EP, the laser can always ``choose" the lower threshold solution for
lasing (upper part of the Riemann sheet), but at the EP only one
solution is available, which is a compromise between the low and high
threshold solution. This gives rise to a local minimum in the
effective gain and inhibits lasing in the vicinity of the EP. In this
way, the essential physics of an EP directly translates into the
physical effects observed in \fig{1}.

\begin{figure}
  \centering
  \includegraphics{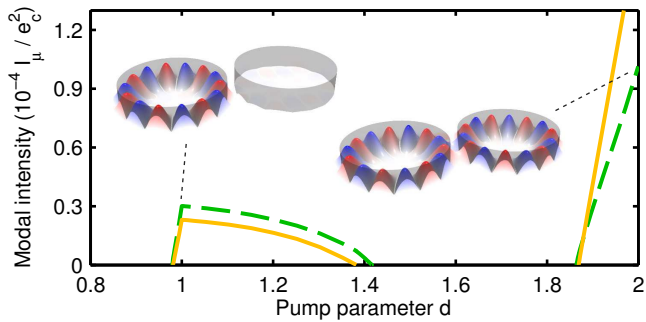}
  \caption{(color online) Laser light intensity emitted from two
    coupled microdisks (see insets) as a function of pump parameter
    $d$. With the same pumping scheme as in \fig{1} (maximum pump in
    each disk: $D=0.71$) the laser is also found here to turn off for
    increasing $d$. The disks have a radius of $45\,{\mu}m$, an
    intercavity spacing of $10\,{\mu}m$ and an index of refraction $n
    =3.67 + 0.09i$.  The gain curve parameters are $k_a\sim2.82\,$THz
    and $\gamma_\bot\sim0.29\,$THz.  The insets show the absolute
    value of the electric field of the green (dashed) lasing mode at
    the pump points $d = 1$ and $d=2$.}
  \label{fig:4}
\end{figure}

To confirm that such a behavior is generic for coupled microlasers, we
consider a device consisting of two coupled circular microdisks
\cite{PreSchMal2008,RamKotShu2011} which has recently been implemented
as an electrically controllable ``photonic-molecule laser''
\cite{FasDeuBen2009}. For this more challenging computation, we have
incorporated in the SALT an advanced higher-order finite element
method discretization technique \cite{*[{We employ the open source
    finite element mesher netgen (}] [{) and the corresponding solver
    NGsolve available from \url{http://sf.net/projects/ngsolve} }]
  Sch1997}. When applying the same pumping scheme as in \fig{1} to
such a coupled microdisk laser (see insets of \fig{4}), we again find
that the laser may turn off when increasing the overall pump power
(see main panel of \fig{4}). Note, however, that more than one mode is
lasing both below and above the turn-off due to a quasidegeneracy of
modes. This example demonstrates that the effect of EPs on the
above-threshold characteristics of a laser should be generally
observable in coupled microlasers.

The success of a corresponding experiment hinges on striking the right
balance between the coupling strength and the amount of gain or loss
in each disk. Gain or loss is fixed by the cavity material as well as
by the applied pumping strength. Microlasers built from active quantum
cascade structures (QCLs) seem to be ideally suited to test this
effect, since they can be well-controlled in their shape and hence in
their coupling. With the active medium being situated within a lossy
metal waveguide \cite{MarDarDeu2011}, a central requirement for
observing this effect, \ie, that the absorption length in the passive
cavity is smaller than the round trip length, has already been
experimentally realized for a variant of such lasers
\cite{SchMujYao2010,*MujSchHof2008}. Also, a sufficiently strong
inter-cavity coupling has been demonstrated for such coupled QCLs
\cite{FasDeuBen2009,SchMujYao2010,*MujSchHof2008}, suggesting that an
experimental realization might be within reach. Note that we have used
realistic material parameters from the experiment \cite{FasDeuBen2009}
in the employed MB equations which typically describe QCLs very well
\cite{WojYuCap2011}.

To summarize, we have shown that the presence of an exceptional point
in the lasing equations for coupled microlasers may be ``brought to
light'' by a suitable variation of the applied pumping. We find that,
in the vicinity of the exceptional point, the output intensity of the
emitted laser light is \emph{reduced}, although the applied pump power
is \emph{increased}. This unorthodox and robust lasing effect may even
turn off a laser above its lowest threshold.  We suggest an
experimental realization of this effect, which would open up many more
possibilities to study the rich physics associated with EPs in the
pump dependence of a laser.

\begin{acknowledgments}
  The authors would like to thank the following colleagues for very
  fruitful discussions: A.~Benz, M.~Brandstetter, C.~Deutsch,
  S.~Esterhazy, G.~Fasching, T.~F\"uhrer, M.~Janits, K.~G.~Makris,
  M.~Martl, J.~M.~Melenk, H.~Ritsch, J.~Sch\"oberl, K.~Unterrainer and
  J.~Vilsecker. Financial support by the Vienna Science and Technology
  Fund (WWTF) through project MA09-030 and by the Austrian Science
  Fund (FWF) through Project No. SFB IR-ON F25-14 as well as
  computational resources by the Vienna Scientific Cluster (VSC) and
  the Yale High Performance Computing Cluster (Yale HPC) are
  gratefully acknowledged. H.E.T.~acknowledges support from the Swiss
  NSF Grant No.~PP00P2-123519/1, NSF Grant No.~EEC-0540832 (MIRTHE),
  and DARPA Grant No.~N66001-11-1-4162 and A.D.S.~from NSF Grant
  No.~DMR-0908437.
\end{acknowledgments}

\end{document}